\documentclass[conference]{IEEEtran}
\IEEEoverridecommandlockouts

\usepackage{graphicx}
\usepackage[bookmarks=false]{hyperref}
\hypersetup{
    colorlinks = true, 
    linkcolor=black, 
    citecolor=black, 
    urlcolor=black 
}
\usepackage{verbatim}
\usepackage{pbox}
\usepackage{pgfplots}
\pgfplotsset{compat=1.16}
\usepgfplotslibrary{fillbetween}
\usepackage{amsmath,amssymb,amsfonts}
\usepackage{listings}
\usepackage{subcaption}

\def\BibTeX{{\rm B\kern-.05em{\sc i\kern-.025em b}\kern-.08em
    T\kern-.1667em\lower.7ex\hbox{E}\kern-.125emX}}

\newtheorem{definition}{Definition}

\begin{document}

\title{Improving Mobile User Interface Testing with Model Driven Monkey Search
    \thanks{This work was supported, in part, by the Science Foundation Ireland grants 13/RC/2094\_P2 and 17/RC-PhD/3485. Paolo Arcaini is supported by ERATO HASUO Metamathematics for Systems Design Project (No. JPMJER1603), JST; Funding Reference number: 10.13039/501100009024 ERATO.}
}

\author{
    \IEEEauthorblockN{Jordan Doyle\IEEEauthorrefmark{1}, Takfarinas Saber\IEEEauthorrefmark{1}, Paolo Arcaini\IEEEauthorrefmark{2}, and Anthony Ventresque\IEEEauthorrefmark{1}}
    \IEEEauthorblockA{\IEEEauthorrefmark{1}Lero@UCD, School of Computer Science, University College Dublin, Dublin, Ireland\\ Email: jordan.doyle@ucdconnect.ie, \{takfarinas.saber, anthony.ventresque\}@ucd.ie}
    \IEEEauthorblockA{\IEEEauthorrefmark{2}National Institute of Informatics, Tokyo, Japan\\Email:     arcaini@nii.ac.jp}
}

\maketitle

\begin{abstract}
Testing mobile applications often relies on tools, such as Exerciser Monkey for Android systems, that simulate user input. Exerciser Monkey, for example, generates random events (e.g., touches, gestures, navigational keys) that give developers a sense of what their application will do when deployed on real mobile phones with real users interacting with it. These tools, however, have no knowledge of the underlying applications' structures and only interact with them randomly or in a predefined manner (e.g., if developers designed scenarios, a labour-intensive task) -- making them slow and poor at finding bugs.

In this paper, we propose a novel control flow structure able to represent the code of Android applications, including all the interactive elements. We show that our structure can increase the effectiveness (higher coverage) and efficiency (removing duplicate/redundant tests) of the Exerciser Monkey by giving it knowledge of the test environment. We compare the interface coverage achieved by the Exerciser Monkey with our new Monkey++ using a depth first search of our control flow structure and show that while the random nature of Exerciser Monkey creates slow test suites of poor coverage, the test suite created by a depth first search is one order of magnitude faster and achieves full coverage of the user interaction elements. We believe this research will lead to a more effective and efficient Exerciser Monkey, as well as better targeted search based techniques for automated Android testing. 
\end{abstract}


\section{Introduction}
Android has become the leading mobile operating system with 72.48\% market share in December of 2020~\cite{androidMarketShare}. The number of mobile app downloads each year has also increased steadily. In 2019, there were 204 billion app downloads, and \$120 billion in revenue generated~\cite{mobileMonetization}. The validation of app quality is now essential, for both keeping users as well as gaining new ones, as app quality is an important factor of user retention~\cite{zuniga2019tortoise}. Ensuring app quality (mostly through testing) is a labour and skill-intensive activity that is often done manually~\cite{kong2018automated,joorabchi2013real} (e.g., developers writing test scenarios) but would benefit from automatic tools~\cite{linares2017continuous}.

However, despite the massive growth in research for automated testing techniques, it has been difficult to create a robust and efficient solution. It is the interactive and event-driven nature of the platform that presents the biggest challenge~\cite{yang2015static}. A user's interaction with an application is highly reliant on their environment. Although mobile interfaces are structured, callback methods are invoked by the Android framework based on the current system state -- which makes the creation of a representation for Android's applications difficult e.g. Control Flow Graphs or Abstract Syntax Trees.

Many different methods of automatic testing have emerged over the years to tackle the issues related to the interactive nature of Android. Some of these methods include automation frameworks~\cite{developers2015monkeyrunner, milano2016androidviewclient, developers2014uiautomator, developers2016espresso, verma2017mobile, zadgaonkar2013robotium}, which provide API's for testers to write interactions as a script, or record and replay tools~\cite{fazzini2017barista, gomez2013reran, halpern2015mosaic, developer2016espressorecorder} that allow developers to record an interaction to be replayed later. While these methods have become popular in the Android community and a standard for Android testing, they are, unfortunately, highly manual and not efficient for large scale development. The majority of research toward Android testing has been devoted to generating user input automatically by means of random~\cite{developers2012ui, machiry2013dynodroid}, systematic~\cite{azim2013targeted, moran2017crashscope, amalfitano2012using}, or model-based~\cite{choi2013guided, su2017guided, amalfitano2014mobiguitar} input generation, removing the manual labour usually required. The most popular and widely used automated framework has been Exerciser Monkey (also known as Monkey)~\cite{developers2012ui}, which is packaged with the Android SDK.

In this paper, we present a novel control flow structure able to represent the code of Android applications, including all the interactive elements. We show that the (control flow) graph generated by our solution can direct/support the execution of a tool like Monkey (We call our extension of Monkey \emph{Monkey++}) and address two of the main challenges in testing Android apps using the framework: removing duplicate and redundant interactions (i.e., efficiency, using less resources) by tracking areas already explored by the test suite; and increasing test coverage (i.e., its effectiveness) by targeting unexplored areas of the application.

Using our graph, we show the coverage achieved by both Monkey, the de facto standard automatic testing tool, and our new Monkey++ using a depth first search. Using 3 open source Android applications, we found that although Monkey achieves at least an average of 85\% interface coverage, it takes 500 interaction attempts to achieve, with only an average of 8\% actually interacting with the application. While the random nature of Monkey creates slow test suites of poor coverage, our search with Monkey++ is one order of magnitude faster and achieves full interface coverage. This shows that even though Monkey can be effective in discovering bugs, it can be made more efficient by integrating knowledge of the application structure enabling better test generation algorithms.

This paper starts by discussing the related work in the area of Android Testing in Section~\ref{Section:Related_Work}. Section~\ref{Section:Control_Flow_Structure} provides a formal definition and implementation of our control flow graph, including its structure, content, and functionality. Section~\ref{Section:Coverage_Metrics} defines the coverage metric we use to evaluate both Monkey and our new Monkey++. Section~\ref{Section:Monkey++} describes our Monkey++ search method. Section~\ref{Section:Evaluation} discusses the results obtained from the Monkey's random search and our Monkey++ DFS. Finally, Section~\ref{Section:Conclusion} concludes the paper.

\section{Related Work}
\label{Section:Related_Work}
Over the years many tools and frameworks have emerged to try improve Android interactive testing and 3 main categories have formed: automation frameworks and APIs, record and replay tools, and automated input generation tools~\cite{linares2017continuous, amalfitano2017general}. While they do succeed in testing Android applications to different degrees, they all have drawbacks preventing them from solving the entire problem. It is important to point out that the set of frameworks discussed in this paper is not exhaustive, but represents the most well-known Android testing frameworks and tools.

\subsection{Automation Frameworks and API's}

Frameworks and API's such as Monkey Runner~\cite{developers2015monkeyrunner}, Android View Client~\cite{milano2016androidviewclient}, UI Automator~\cite{developers2014uiautomator} as well as many others~\cite{developers2016espresso, verma2017mobile, zadgaonkar2013robotium}, allow developers to create testing scripts that run interactive events through the Android emulator. These tools allow users to test their application using the user interface in a similar manner to unit testing, which provides functional testing of application code as well as testing runtime behaviour of the activity lifecycle and user input callback methods. The developer is required to code an interaction with the device and then test whether the resulting outcome is correct, by either image comparison, view attribute comparison or by simply monitoring the results of the test manually. This type of testing is useful for basic functional testing during the development process as the testing environment is highly controlled and stable. However, these tools require hours of manual scripting to be maintained as the application under test (AUT) develops, and scripts are not guaranteed to work on all devices, due to Android device fragmentation~\cite{linares2017continuous}. Record and replay tools~\cite{fazzini2017barista, gomez2013reran, halpern2015mosaic, developer2016espressorecorder} such as Mosaic or Espresso Recorder provide the same features as automation API's but they eliminate the need for hours of scripting. Instead of scripting interactions with an Android application, a developer simply uses their application on an emulator or device and the interactions are recorded. These interactions can then be replayed whenever the developer needs them. This method provides a quicker way of creating tests but does not solve the core problems. The recordings are still time consuming and reliant on the device used to create them. 

\subsection{Automated Input Generators}

The most active area in mobile software testing is currently automated input generation. Three of the most used methods are random, systematic, and model-based input generation~\cite{linares2017continuous, wang2018empirical}.

Using \textbf{random input} for testing an application's interface is the simplest of methods but they are generally very inefficient and lead to a high number of redundant and repetitive interactions. An example of this is Monkey, a program that runs on your emulator or device and generates pseudo-random streams of user events such as clicks, touches, or gestures, as well as a number of system-level events~\cite{developers2012ui}. In practice, Monkey does not support keyboard input or system notifications, and most of the random inputs applied to the interface do nothing within the application. Dynodroid, another well-known random input generator tries to overcome this issue by applying a Frequency Strategy and Biased-Random Strategy, to increase the efficiency of the inputs it selects. The frequency strategy uses the inputs that are most frequently used and the biased-random strategy uses the inputs that are relevant in most contexts~\cite{machiry2013dynodroid}. Unlike Monkey, Dynodroid also supports keyboard input and system notifications, however, this is possibly its downfall. For instance, in order to provide system events, Dynodroid needed to instrument the Android framework~\cite{machiry2013dynodroid} and in doing this they made Dynodroid hard to update and maintain which has led to the framework becoming outdated and unused.

Despite Dynodroid performing better than Monkey, showing an increased application coverage and less interaction redundancy, Monkey has remained one of the most popular frameworks available for automated interface testing. This could be attributed to Monkey being maintained and packaged as part of the Android framework and its simple implementation, resulting in the majority of publishers comparing their results with results from Monkey.

\textbf{Systematic input} involves dynamically analysing the interface of an application and generating appropriate test inputs as they are found. This method provides effective testing of an application without prior knowledge of the interface structure or the underlying code. A good example of this technique is Android Ripper that maintains a state machine model of the GUI, called a GUI Tree. The GUI Tree model contains the set of GUI states and state transitions encountered during the ripping process~\cite{amalfitano2012using}. Similar to Android Ripper, tools such as Automatic Android App Explorer (A3E) and Crash Scope use the same systematic approach to exploring an application interface, however, they also use static analysis to determine activity or method interactions and contextual features respectively~\cite{azim2013targeted, moran2017crashscope}. 

A systematic approach to automated input generation shows promise with Android Ripper, A3E and CrashScope showing varying degrees of improvement over Monkey in areas such as coverage, input redundancy and bug detection~\cite{wang2018empirical, amalfitano2012using, azim2013targeted, moran2017crashscope}, however, a systematic approach can never guarantee complete coverage of the application.

\textbf{Model-based} testing requires a formal model of the application, such as an event flow graph, control flow graph, finite state machine, etc. Once the model has been made, it is used to generate a test suite of device inputs. For example, Stoat~\cite{su2017guided} and MobiGuitar~\cite{amalfitano2014mobiguitar} use a finite state machine (FSM) to model the AUT. Stoat uses dynamic analysis, enhanced by a weighted UI exploration strategy and static analysis, to explore the app’s behaviours and construct a stochastic FSM~\cite{su2017guided}. MobiGuitar on the other hand, uses an enhanced version of Android Ripper and dynamically traverses the application in a breadth first fashion to generate its FSM~\cite{amalfitano2014mobiguitar}.

Model-based approaches to automated Android testing hold up well against the other testing methods and tools we have discussed but have failed to generate a solid user base. Stoat sets itself apart with its method of using system events within tests. Instead of trying to model system events, it randomly injects various system-level events into its UI tests~\cite{su2017guided}. This simulates the random nature of receiving notifications and state changes in a real environment. However, Stoat is also ineffective with regular gestures (e.g., Pinch Zoom, Move) and specific input data formats~\cite{su2017guided}, while MobiGuitar's reliance on Android Ripper caused it to have the same pitfalls.

\section{Control Flow Structure}
\label{Section:Control_Flow_Structure}
In this section we define the data structure (control flow graph) we extract from Android applications and how it is implemented.

\subsection{Formal Definition}\label{Subsection:Formal_Def}

\vspace{1mm}
\begin{definition}[Control Flow Structure]
A \emph{Control Flow Structure} is a directed graph $G=(V,E)$ where $V=\{v_1,$ $\ldots,$ $v_n\}$, $n\in \mathbb{N}$, is a set of vertices and $E=\{e_1, \ldots, e_m\}$, $m\in \mathbb{N}$, is a set of arcs (directed edges), where $e_i=(v_j,v_k)$ with $v_j,v_k \in V$.
\end{definition}
\vspace{2mm}

Our control flow graph $G$ is composed of three types of vertices: $V=V_s\cup V_m\cup V_{\mathit{ui}}$, that represent the three important elements of a mobile application:
\begin{itemize}
\item $V_s$ represents the statement vertices. While testing, we want to focus on the code written for the application rather than system or library code, so we exclude them from our statement set. Statements play an important role by revealing application logic, internal method calls and the overall control flow of the application.
\item $V_m$ represents the method vertices and similarly only comprises methods developed for the application. The method vertices represent the entry points to a method and lead to the statement vertices within. These vertices can be further split into three types: the Android life-cycle methods, the user input callback methods, and standard Java methods.
\item $V_{\mathit{ui}}$ represents the interface vertices, indicating user input locations within the graph.
\end{itemize}

Not all control flow edges can be included in our graph as they are runtime dependent. For example, Activity lifecycle methods are called by the Android framework based on the current state of the system or application, meaning they have no explicit call within the application code. Therefore, an additional primitive runtime model was created to track and enforce the proper movement between these methods during a search. This model is discussed in Section~\ref{Section:Monkey++}.

\begin{figure}[h]
\includegraphics[width=\linewidth]{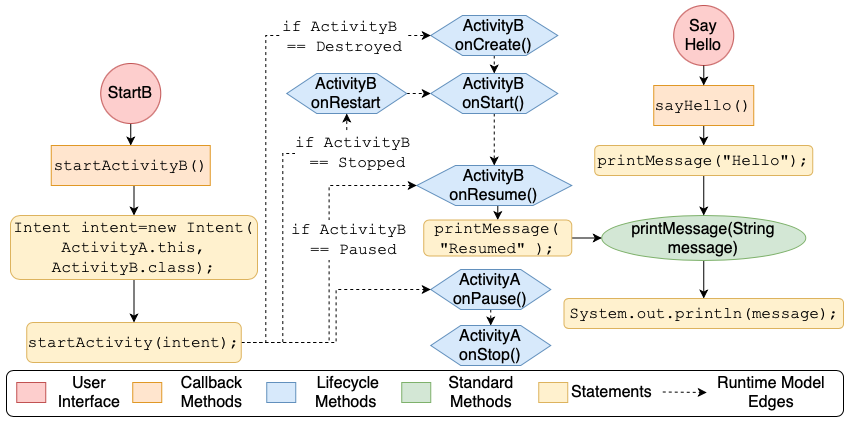}
\caption{Control flow structure representing a small Android application. User interface (e.g., Buttons) are red, callback methods are orange, lifecycle methods are blue, standard methods are green, and statements are yellow.} \label{Figure:Control_Flow_Example}
\end{figure}

Figure~\ref{Figure:Control_Flow_Example} shows an example of a control flow structure representing an Android application: its syntax (methods and statements) and the user interactions it contains. We can see from this example two interface views, 'StartB' and 'SayHello', connected to their respective method listeners 'startActivityB' and 'sayHello'. The statements within these methods are linked by edges in their execution order. Method call edges are also shown, for instance the statement 'startActivity(intent)' calls the subsequent lifecycle methods. Some of the edges are shown as a dashed arrow. These edges are not contained in our graph, instead they are determined by the primitive runtime model mentioned above. It decides which edge to follow during a search based on the status of the called activity. In order to simplify our example, we include statement nodes for the 'ActivityB onResume()' lifecycle method only.

\subsection{Implementation}

Traditional control flow analysis cannot be directly applied to Android applications because they are framework based and event-driven. For this reason we used our own algorithms along with existing frameworks (i.e., Soot and FlowDroid) to statically analyse the application. 

\subsubsection{Soot} started off as a Java optimization framework, by now, researchers and practitioners use Soot to analyse, instrument, optimize and visualize Java and Android applications. Soot's primary functionality is inter-procedural and intra-procedural analysis using an intermediate representation. It takes as input Java source and byte-code, or more recently Android byte-code, and transforms it into an intermediary representation for easier analysis~\cite{lam2011soot}. 

Soot's primary intermediate representation is Jimple---a typed three address code~\cite{lam2011soot}. The creation of Jimple was motivated by the difficulty of directly analysing Java byte-code: although it is possible to construct a control-flow graph for Java byte-code, the implicit stack masks the flow of data, making it difficult to analyse~\cite{vallee1998jimple}. Using Jimple makes the analysis easy due to its low complexity.

Our motivation for choosing Soot is due to the framework and the Jimple intermediate representation being the most adopted support tool and format respectively~\cite{li2017static}, as well as the internal objects that Soot provides to represent the code and Soot's inter-procedural and intra-procedural analysis.

However, it does have its drawbacks: although Soot has the ability to analyse Android byte-code, it does not have knowledge of the activity lifecycle and therefore cannot get the body of lifecycle methods. Also, Android applications have multiple entry points as part of the lifecycle model, this means that Soot cannot create a call graph as it has no knowledge of where to enter the application. To solve this issue we use FlowDroid.

\subsubsection{FlowDroid} is the first fully context-, field-, object- and flow-sensitive taint analysis which considers the Android application lifecycle, while also featuring a novel, particularly precise variant of an on-demand alias analysis~\cite{arzt2014flowdroid}. It was designed to analyse applications, and alert users of malicious data flows, or as a malware detection tool which could determine if a leak was a policy violation.

Despite its main purpose, FlowDroid’s knowledge of the activity lifecycle is why it became applicable to this research. As mentioned, Android applications do not have a main method, they consist of entry points that are called by the Android framework when needed. These entry points can be input callback methods, defined to respond to a user input; or lifecycle methods, defined by the developer to respond to a system event, for example, launching a new activity.

FlowDroid adds the ability for Soot to retrieve the body of lifecycle methods and include them in its analysis. Additionally, it allows Soot to create a call graph by generating a dummy main method using Heros, an IFDS/IDE solver framework which is not path sensitive, but instead joins analysis results immediately at any control flow merge point~\cite{arzt2014flowdroid}. This main method contains every order of individual lifecycle and callback components without traversing all the individual paths.

\subsubsection{The control flow structure} is mostly populated using Soot and FlowDroid's static analysis. However, even though the structures provided by these frameworks are comprehensive and invaluable, they are also disconnected and do not allow for easy traversal. For this reason, we use the data they provide to populate our own graph implementation, consisting of a set of vertex and edge objects. An edge object consists of a source and target vertex, while a vertex object contains several attributes including but not limited to a unique ID, a label, and a type. As discussed in Section~\ref{Subsection:Formal_Def} our control flow structure consists of 3 types of vertices.

\begin{itemize}
\item \textbf{Method} vertices are retrieved from FlowDroid's call graph. We check each method in the call graph to further classify them as activity lifecycle, input callback, or standard Java methods as well as filtering methods from the Java and Android SDK, before adding them and their associated call edges to our graph.
\item \textbf{Statement} vertices are gathered from the UnitGraph objects created by Soot for each method in the analysed application. Each UnitGraph contains a unit chain detailing all the Jimple statements and their execution order within the method. We add each Jimple statement in the unit chain to our graph, while also checking for method calls within the statement and creating the relevant edges. 
\item \textbf{Interface} controls and callback methods are identified by FlowDroid and included in the graph. However, FlowDroid fails to provide a link between the individual controls and their associated callback method. This is made more challenging by most callback methods having the same name when defined within the Java code, e.g., onClick(). Currently we instrument the AUT so that the control ID for each callback method can be found in the Jimple statements. Finding better methods of retrieving the interface data is part of future work.
\end{itemize}

As mentioned in Section~\ref{Subsection:Formal_Def}, not all control flow edges can be determined through static analysis, as they are runtime dependent, and therefore cannot be included in the graph. For example, activity lifecycle methods are called by the Android framework based on the current state of the system or application, meaning they have no explicit call within the application code. This causes disconnected clusters to form in the graph. The control flow to and from these clusters is determined at runtime by the Android framework, or by our runtime model (see Section~\ref{Section:Monkey++}) during a search.

\begin{figure}[h]
\centering\includegraphics[width=50mm]{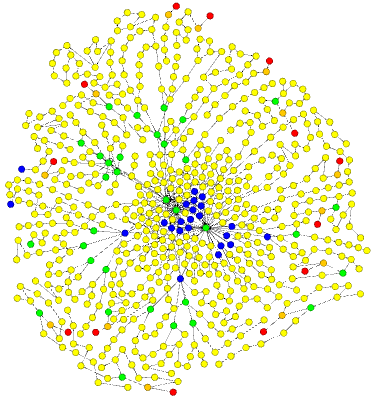}
\caption{Control flow structure for Activity Lifecycle application. User interface (e.g., Buttons) are red, callback methods are orange, lifecycle methods are blue, standard methods are green, and statements are yellow.}
\label{Figure:Control_Flow_Graph_Example}
\end{figure}

Figure~\ref{Figure:Control_Flow_Graph_Example} shows the control flow structure generated for one of our AUT, Activity Lifecycle. The visualisation of the graph was generated using the DOT visualisation language and Gephi.

\section{Coverage Metrics based on our Control Flow}
\label{Section:Coverage_Metrics}
Once we have a representation of the Android app (i.e., the graph $G$ we have proposed in Section~\ref{Section:Control_Flow_Structure}), we can look at the code coverage of tests generated by different test generation techniques, i.e., how much of the graph is covered by the test suites generated by a given technique.

\subsection{Formal Definitions}

In the current paper, we are interested in the interactive elements (e.g., buttons) of the application. The focus on interactive elements (i.e., $v_j\in V_{\mathit{ui}}$ in our graph) tells us how much of what the users can do is tested. Further research will involve using the methods $V_{m}$ and statements $V_{s}$ to enhance our search to target specific features likely to uncover application bugs.

Let's first introduce the notion of test and test generation tool.

\vspace{2mm}
\begin{definition}[Test and test suite]\label{Definition2:Test}
Let us assume a mobile application $\mathcal{A}$ and the graph representing it $G=(V,E)$. A \emph{test} is a sequence of vertices $t=[v^t_1, v^t_2, \ldots, v^t_m]$ such that $(v^t_i, v^t_{i+1}) \in E$ (for $i = 1, 
\ldots, m-1$), and $v^t_1$ is an interface vertex (i.e., $v^t_1 \in V_{\mathit{ui}}$). A test suite is a set of tests $\mathit{TS} = \{t_1, \ldots, t_k\}$.
\end{definition}
\vspace{2mm}

\begin{definition}[Test generation tool]
Given a mobile application $\mathcal{A}$ and its graph $G=(V,E)$, a \emph{test generation tool} $\mathcal{T}$ can be seen as a function producing a test suite $\mathit{TS}$, i.e., $\mathit{TS} = \mathcal{T}(\mathcal{A}, G)$.
\end{definition}
\vspace{2mm}

As said before, in our work we are interested in measuring how much of the interface of the application has been covered. Therefore, we introduce the notion of interface coverage as follows.

\vspace{2mm}
\begin{definition}[Interface Coverage]
Let us assume a mobile application $\mathcal{A}$, its graph $G=(V,E)$ (with $V=V_s\cup V_m\cup V_{\mathit{ui}}$), and a test generation tool $\mathcal{T}$. The \emph{interface coverage} achieved by the tool is defined as:
\[\mathit{IC}(\mathcal{T}, \mathcal{A}, G)=\frac{|\bigcup_{t \in \mathit{TS}} \{v \in t \mid v \in V_{\mathit{ui}}\}|}{|V_{\mathit{ui}}|}\]
where $\mathit{TS} = \mathcal{T}(\mathcal{A}, G)$.
\end{definition}
\vspace{2mm}

The interface coverage tells us how many of the interactive elements in the whole application are explicitly being tested.

\begin{figure}
\includegraphics[width=\linewidth]{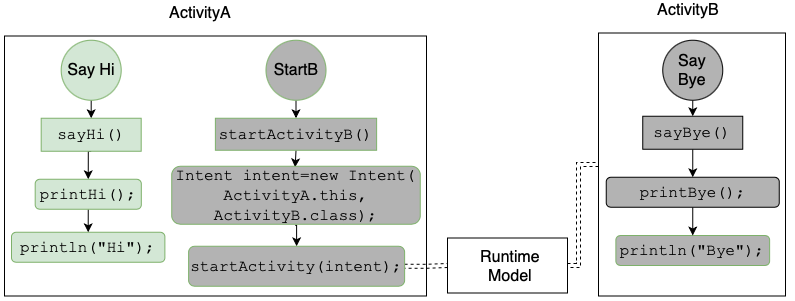}
\caption{Interaction Scenario: When ``SayHi'' is touched we achieve 33\% interface coverage but, to achieve 100\% interface coverage, we must touch ``StartB'' for the search to reach ``SayBye''.}\label{Figure:Coverage_Metrics_Example}
\end{figure}

Figure~\ref{Figure:Coverage_Metrics_Example} shows an interaction scenario (i.e., a test, see definition~\ref{Definition2:Test}), where green nodes identify covered vertices. While interface elements are source vertices in the graph they are not entry points within the Android application. We can see that when ``SayHi'' is touched we achieve 33\% interface coverage but, in order to achieve 100\% interface coverage, we must touch ``StartB'' in ``ActivityA'' (assuming ``ActivityA'' is an entry point class), so that the search can reach ``SayBye'' in ``ActivityB''.

\subsection{Coverage Evaluation}\label{Subsection:Coverage_Evaluation}

In order to obtain the coverage achieved by the Monkey test suite (sequence of interaction events), we needed to map it to our control flow graph. We first needed to extract it from Monkey's output, which is limited to the command line. We ran Monkey as a Python sub-process and captured the output produced, which contains a stream of events executed on the device during a test. A sample of the output can be seen in Listing~\ref{Code:Monkey_Output} showing the generated random interactions (instructions proceeded by ':') that were executed on an Android device.

\vspace{2mm}
\begin{lstlisting}[language={}, label={Code:Monkey_Output}, abovecaptionskip=5pt, caption={Sample of output given from Monkey}]
:Switch: #Intent;action=android.intent.action.MAIN;category=android.intent.category.LAUNCHER;launchFlags=0x10200000;component=com.example.android.lifecycle/.ActivityA;end
// Allowing start of Intent { act=android.intent.action.MAIN cat=[android.intent.category.LAUNCHER] cmp=com.example.android.lifecycle/.ActivityA } in package com.example.android.lifecycle
:Sending Touch (ACTION_DOWN): 0:(177.0,1604.0)
:Sending Touch (ACTION_UP): 0:(170.91504,1612.0048)
// Allowing start of Intent { cmp=com.example.android.lifecycle/.DialogActivity } in package com.example.android.lifecycle
:Sending Touch (ACTION_DOWN): 0:(114.0,1359.0)
// activityResuming(com.example.android.lifecycle)
:Sending Touch (ACTION_UP): 0:(112.05528,1361.9899)
\end{lstlisting}
\vspace{2mm}

Regular expressions were used to search the output text and identify the type of events used during a test, as well as any execution details that accompany them. For example, listing~\ref{Code:Extracting_Switch_Event} shows the method used to extract the switch event given on line 1 of listing~\ref{Code:Monkey_Output}.

\vspace{2mm}
\begin{lstlisting}[language=python, label=Code:Extracting_Switch_Event, abovecaptionskip=10pt, caption={Extracting switch events from Monkey output}]
def __extract_switch_event(self, line):
    ptn=re.compile(r':Switch:\s.+;component=([\w./]+);end')
    component=regex_search(ptn, line, 1)
    
    next_line = self.__monkey_output.pop(0)
    if any(word in next_line for word in ["Allowing","Resuming","Rejecting"]) and component in next_line:
        accepted=False if "Rejecting" in next_line else True
    else:
        self.__monkey_output.insert(0, next_line)
        
    self.events.append(SwitchEvent(component, accepted))
\end{lstlisting}
\vspace{2mm}

The Monkey output only provides screen coordinates for the interactions executed. In order to determine which of the AUT's interface control views were used during a test, we needed to convert those coordinates into the control view at that location. UI Automator~\cite{developers2014uiautomator}, an automation framework, has a dump feature that provides a structured layout of views currently on the screen, including each views ID, screen coordinates and whether it is clickable. Due to the dump feature only working on a screen's current content, we were forced to re-execute the extracted events while collecting details of the view found at the coordinates, if one existed. Listing~\ref{Code:Running_Touch_Events} shows a small portion of the code used to execute a touch event.

\vspace{2mm}
\begin{lstlisting}[language=python, label=Code:Running_Touch_Events, abovecaptionskip=5pt, caption={Running Touch Events on the Application}]
window = device.get_focused_window()
if action is not None:
  view_id = device.find_view(action['x'], action['y'])

  if view_id is not None:
    view={'id':view_id,'activity':window,'coords':action}
    
  device.touch(action['x'], action['y'])
\end{lstlisting}
\vspace{2mm}

Once the Monkey test suite had been extracted, the search followed the graph to determine where in the application Monkey travelled and thus gave us the application coverage achieved.

\section{Monkey++: Monkey Driven by our Control Flow}
\label{Section:Monkey++}
Besides defining coverage metrics, our control flow could also empower Monkey in its search for better test suites. Instead of randomly testing the application, we propose Monkey++; a Monkey which is capable of leveraging our control flow using some graph navigation algorithm (e.g., Depth First Search, Breadth First Search, or more elaborate techniques) to achieve a better coverage and with a smaller effort.

\subsection{Monkey++ with Depth First Search}
As a use-case, we developed Monkey++ which navigates our control flow using the Depth First Search algorithm.

Monkey++'s search begins with the default launch activity within the AUT. Monkey++ gets all the interface vertices for an activity and performs a search with one vertex as the root, recursively finding new interface controls while it searches. When all the interface vertices in the current activity have been explored, Monkey++ moves back one activity, the equivalent to pressing the back button on a device. 

As mentioned previously, our graph does not contain control flow edges between activity lifecycle methods as these control flows are determined by the Android framework based on the current system state and the application state itself. Our search determines these flows at runtime by tracking both the activity stack and lifecycle status of each activity.

The stack determines which activity is currently on the screen. For example, when a new activity is launched, it gets added to the top of the stack and takes over the screen, conversely if the back button is pressed, the top activity in the stack is removed and the previous one takes its place.

The lifecycle status of each activity determines which lifecycle methods need to be executed/searched when an activity is added, or removed from the top of the stack. For example, the onCreate() lifecycle method is only executed if an instance of the activity does not already exist in the stack, while the onResume() method is only executed if the activity is moved to the top of the stack and was previously in a paused state.

When visiting a vertex in our graph we check for 3 features: an activity launch statement; an activity finish statement; or a lifecycle method. If any of these features are found then the current state of the stack and the lifecycle status of each activity will determine where the search continues. This can be seen in listing~\ref{Code:Search_Method} showing the main DFS search method.

\vspace{2mm}
\begin{lstlisting}[language=Java, label=Code:Search_Method, abovecaptionskip=5pt, caption={Search method for executing a DFS on a graph}]
protected void search(Vertex v) {
        v.visit();
        v.localVisit();

        for(DefaultEdge e : graph.outgoingEdgesOf(v)) {
            Vertex target = graph.getEdgeTarget(e);
            if(!target.hasLocalVisit()) {
                search(target);
            }
        }

        switch(v.getType()) {
            case statement:
                checkForIntentAndStart(v);
                checkForActivityFinish(v);
                break;
            case lifecycle:
                updateLifecycle(v);
                break;
        }
    }
\end{lstlisting}
\vspace{2mm}

While searching the graph we enter and exit activities repeatedly, meaning certain lifecycle methods need to be executed/searched multiple times to maintain the activity stack and lifecycle states. In a standard DFS, repetition is not allowed as it leads to searching the same vertices infinitely. We allow repetition in our overall search by creating smaller local searches that do not allow repetition. While each local search cannot visit the same vertex more than once, the area covered by each local search can overlap.

\section{Validation}
\label{Section:Evaluation}
The Monkey is capable of finding bugs within Android applications but its performance can be greatly improved by adding knowledge of the AUTs underlying structure. Leveraging knowledge of the internal structure using even a simple search method such as DFS can provide a more efficient and effective Monkey. We demonstrate this by comparing the coverage achieved by a standard Monkey random search with a DFS using our (control flow) graph (Monkey++).

Monkey supports different interaction types, such as screen touches, key presses, swipe events, etc. In order to get the best possible coverage from Monkey, we limit the interactions to screen touches, as these are the only interactions that the AUTs accept. For both Monkey and Monkey++ a test is a sequence of interactions where an interaction is a click on the screen. Monkey, with no knowledge of the interface, can not be specific, so an interaction does not always touch an interface control. Conversely, thanks to the (control flow) graph, every Monkey++ interaction is specific to an interface control (Similarly to getting interface controls views from Monkey coordinates in Section~\ref{Subsection:Coverage_Evaluation}, we can also get coordinates of interface controls views from control IDs).    

Stress testing the application was not the objective, therefore we gave Monkey a 0.5 second throttle to allow plenty of time between events. For both Monkey and Monkey++ our validation centres around \emph{interface coverage}. 

Due to Monkey executing random events on the device, all experiments are repeated 10 times executing 500 device interactions each. Results for Monkey are averaged over the 10 obtained test suites. However, using the DFS as the underlying graph navigation for Monkey++ makes it deterministic--thus only requiring one run.

\subsection{Applications Under Test}

All the AUTs are free and open source applications taken from Fossdroid~\cite{fossdroid}. The applications were chosen randomly with the following criteria in mind: relatively small in size; and simple (i.e., no custom views, and minimal external libraries). These criteria were included due to the immature nature of our graph generation. Further development is needed to allow for larger industrial applications. The chosen applications can be seen in Table~\ref{Table:Composition_of_AUTs}.

\begin{table}[!tb]
\caption{Composition of AUTs}
\begin{center}
\begin{tabular}{|c|c|c|c|}
\hline
\textbf{App Name} & \textbf{No. Activities} & \textbf{No. Controls} & \textbf{Java LOC} \\
\hline
Mo Clock & 3 & 10 & 378 \\
\hline
Activity Lifecycle & 4 & 13 & 558 \\
\hline
Volume Control & 4 & 11 & 2254 \\
\hline
\end{tabular}
\label{Table:Composition_of_AUTs}
\end{center}
\end{table}

\subsection{Interface Coverage}

\begin{figure}[!tb]
\centering
\begin{subfigure}[t]{1\columnwidth}
\begin{tikzpicture}
\begin{semilogxaxis}[
    height=5.2cm,
    width=\linewidth,
    legend pos=north west,
    xmajorgrids=true,
    ymajorgrids=true,
    grid style=dashed,
    xlabel={No. Interactions},
    ylabel={Interface Coverage (\%)},
    xmin=1, xmax=500,
    ymin=0, ymax=100,
    log ticks with fixed point,
    xtick={1,10,100,200,500},
    ytick={0,20,40,60,80,100},
    legend image post style={scale=0.2},
    legend style={
        font=\scriptsize,
        legend cell align=left
    }
]
\legend{Monkey++,,Monkey Avg,,Monkey max/min,}
\addplot[color=red]
coordinates {(1,0)(2,10)(3,20)(4,30)(5,40)(6,50)(7,60)(8,60)(9,70)(10,70)(11,70)(12,80)(13,80)(14,80)(15,80)(16,90)(17,90)(18,90)(19,100)};
\addplot[color=gray,name path=max]
coordinates {(1,0)(2,0)(3,0)(4,10)(5,10)(6,10)(7,10)(8,20)(9,20)(10,20)(15,30)(20,30)(25,30)(30,40)(35,40)(40,40)(45,40)(50,40)(60,50)(70,50)(80,50)(90,50)(100,60)(115,60)(130,70)(145,70)(160,80)(180,80)(200,80)(225,90)(250,90)(275,90)(300,100)(350,100)(400,100)(450,100)(500,100)};
\addplot[black, thick ,name path=ave]
coordinates {(1,0)(2,0)(3,0)(4,0)(5,0)(6,0)(7,4)(8,4)(9,6)(10,6)(15,14)(20,14)(25,16)(30,17)(35,20)(40,24)(45,25)(50,27)(60,29)(70,32)(80,38)(90,38)(100,38)(115,40)(130,45)(145,46)(160,49)(180,58)(200,64)(225,68)(250,73)(275,78)(300,80)(350,80)(400,83)(450,84)(500,88)};
\addplot[color=gray,name path=min]
coordinates {(1,0)(2,0)(3,0)(4,0)(5,0)(6,0)(7,0)(8,0)(9,0)(10,0)(15,10)(20,10)(25,10)(30,10)(35,10)(40,10)(45,10)(50,10)(60,10)(70,20)(80,20)(90,30)(100,30)(115,30)(130,30)(145,30)(160,30)(180,30)(200,30)(225,40)(250,40)(275,50)(300,50)(350,60)(400,60)(450,70)(500,70)};
\addplot[gray!60] fill between[of=max and ave];
\addplot[gray!60] fill between[of=min and ave];
\end{semilogxaxis}
\end{tikzpicture}
\caption{Mo Clock}
\label{Mo_Clock:Interface_Coverage}
\end{subfigure}

\begin{subfigure}[t]{1\columnwidth}
\begin{tikzpicture}
\begin{semilogxaxis}[
    height=5.2cm,
    width=\linewidth,
    legend pos=north west,
    xmajorgrids=true,
    ymajorgrids=true,
    grid style=dashed,
    xlabel={No. Interactions},
    ylabel={Interface Coverage (\%)},
    xmin=1, xmax=500,
    ymin=0, ymax=100,
    log ticks with fixed point,
    xtick={1,10,100,200,500},
    ytick={0,20,40,60,80,100},
    legend image post style={scale=0.2},
    legend style={
        font=\scriptsize,
        legend cell align=left
    }
]
\legend{Monkey++,,Monkey Avg,,Monkey Max/Min,}
\addplot[color=red]
coordinates {(1,0)(2,8)(3,15)(4,23)(5,31)(6,38)(7,46)(8,54)(9,62)(10,69)(11,77)(12,85)(13,85)(14,92)(15,92)(16,92)(17,100)};
\addplot[color=gray,name path=max]
coordinates {(1,0)(2,8)(3,8)(4,8)(5,15)(6,15)(7,15)(8,15)(9,15)(10,15)(15,15)(20,23)(25,23)(30,31)(35,31)(40,31)(45,31)(50,31)(60,38)(70,46)(80,46)(90,54)(100,54)(115,62)(130,62)(145,69)(160,69)(180,69)(200,77)(225,77)(250,85)(275,85)(300,85)(350,92)(400,100)(450,100)(500,100)};
\addplot[black, thick ,name path=ave]
coordinates {(1,0)(2,1)(3,1)(4,3)(5,4)(6,6)(7,6)(8,7)(9,9)(10,9)(15,10)(20,12)(25,15)(30,19)(35,22)(40,22)(45,22)(50,25)(60,30)(70,31)(80,36)(90,40)(100,44)(115,48)(130,48)(145,53)(160,56)(180,58)(200,64)(225,68)(250,73)(275,76)(300,79)(350,82)(400,83)(450,84)(500,85)};
\addplot[color=gray,name path=min]
coordinates {(1,0)(2,0)(3,0)(4,0)(5,0)(6,0)(7,0)(8,0)(9,8)(10,8)(15,8)(20,8)(25,8)(30,8)(35,8)(40,8)(45,8)(50,15)(60,15)(70,15)(80,23)(90,31)(100,31)(115,31)(130,31)(145,38)(160,38)(180,46)(200,46)(225,54)(250,54)(275,62)(300,69)(350,77)(400,77)(450,77)(500,77)};
\addplot[gray!60] fill between[of=max and ave];
\addplot[gray!60] fill between[of=min and ave];
\end{semilogxaxis}
\end{tikzpicture}
\caption{Activity Lifecycle}
\label{Activity_Lifecycle:Interface_Coverage}
\end{subfigure}

\begin{subfigure}[t]{1\columnwidth}
\begin{tikzpicture}
\begin{semilogxaxis}[
    height=5.2cm,
    width=\linewidth,
    legend pos=north west,
    xmajorgrids=true,
    ymajorgrids=true,
    grid style=dashed,
    xlabel={No. Interactions},
    ylabel={Interface Coverage (\%)},
    xmin=1, xmax=500,
    ymin=0, ymax=100,
    log ticks with fixed point,
    xtick={1,10,100,200,500},
    ytick={0,20,40,60,80,100},
    legend image post style={scale=0.2},
    legend style={
        font=\scriptsize,
        legend cell align=left
    }
]
\legend{Monkey++,,Monkey Avg,,Monkey max/min,}
\addplot[color=red]
coordinates {(1,0)(2,9)(3,18)(4,27)(5,36)(6,45)(7,54)(8,63)(9,72)(10,72)(11,81)(12,81)(13,100)};
\addplot[color=gray,name path=max]
coordinates {(1,0)(2,9)(3,9)(4,9)(5,18)(6,18)(7,18)(8,27)(9,27)(10,36)(15,36)(20,45)(25,45)(30,54)(35,54)(40,63)(45,63)(50,72)(60,72)(70,72)(80,72)(90,72)(100,81)(115,81)(130,90)(145,90)(160,90)(180,100)};
\addplot[black, thick ,name path=ave]
coordinates {(1,0)(2,3)(3,3)(4,5)(5,5)(6,10)(7,10)(8,15)(9,20)(10,30)(15,32)(20,39)(25,40)(30,42)(35,45)(40,52)(45,56)(50,60)(60,65)(70,66)(80,66)(90,68)(100,69)(115,72)(130,75)(145,76)(160,80)(180,86)(200,88)(225,90)(250,91)(275,92)(300,95)(350,100)(400,100)(450,100)(500,100)};
\addplot[color=gray,name path=min]
coordinates {(1,0)(2,0)(3,0)(4,0)(5,0)(6,4)(7,4)(8,7)(9,10)(10,15)(15,22)(20,29)(25,29)(30,35)(35,39)(40,43)(45,48)(50,52)(60,52)(70,52)(80,61)(90,62)(100,62)(115,65)(130,65)(145,67)(160,70)(180,73)(200,75)(225,76)(250,76)(275,78)(300,81)(350,85)(400,88)(450,89)(500,90)};
\addplot[gray!60] fill between[of=max and ave];
\addplot[gray!60] fill between[of=min and ave];
\end{semilogxaxis}
\end{tikzpicture}
\caption{Volume Control}
\label{Volume_Control:Interface_Coverage}
\end{subfigure}
\caption{Interface Coverage (over 10 runs of Monkey and 1 run of Monkey++). Note the log scale of x-axis.}
\end{figure}
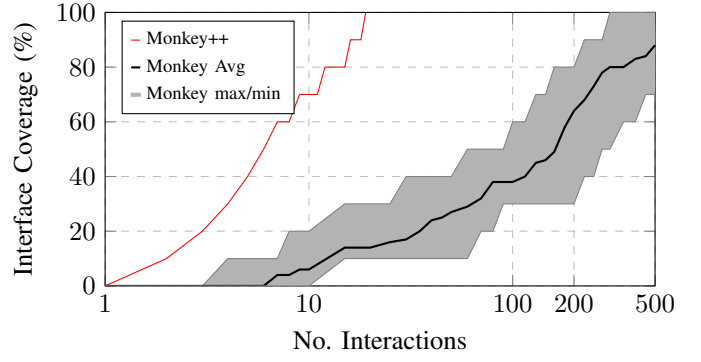
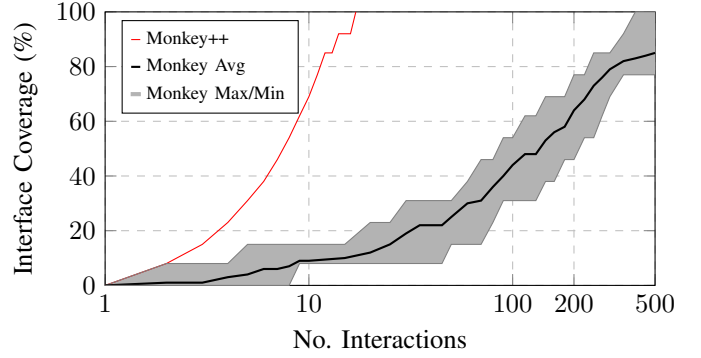
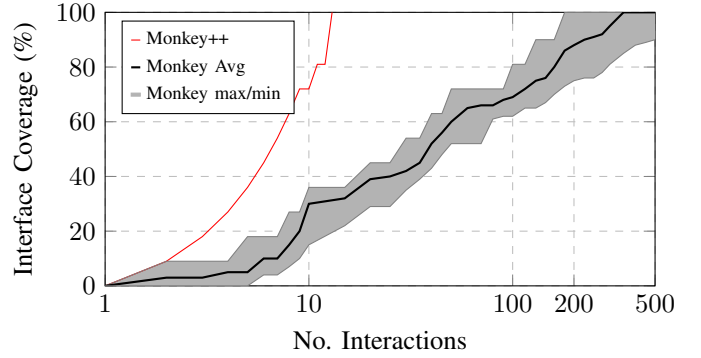

Figures~\ref{Mo_Clock:Interface_Coverage}--\ref{Volume_Control:Interface_Coverage} show the interface coverage achieved by both Monkey and Monkey++ for the three AUTs.

Table~\ref{Table:Results_Overview} shows the coverage achieved by each AUT and the number of interactions required to reach it.

\begin{table}[!tb]
\caption{Overview of Results -- Interface coverage (No. of interactions required to achieve it)}
\begin{center}
\setlength{\tabcolsep}{4pt}
\begin{tabular}{|c|c|c|c|c|}
\hline
 &  & \multicolumn{3}{|c|}{\textbf{Monkey}} \\
\cline{3-5}
\textbf{App Name} & \textbf{Monkey++} & \textbf{Average} & \textbf{Min} & \textbf{Max} \\
\hline
Mo Clock & 100\% (19) & 88\% (500) & 70\% (500) & 100\% (300) \\
\hline
Activity Lifecycle & 100\% (17) & 85\% (500) & 77\% (500) & 100\% (400) \\
\hline
Volume Control & 100\% (13) & 100\% (350) & 90\% (500) & 100\% (180) \\
\hline
\end{tabular}
\label{Table:Results_Overview}
\end{center}
\end{table}

Even though Monkey achieved an at least average of 85\% coverage, it required 500 interaction attempts to do it. We expected with such a low number of interface controls, that the Monkey would get 100\% coverage in less attempts. Even when 100\% coverage was achieved, it required at least 180 interactions. Given an application with a more complex interface structure, we would expect Monkey to achieve a much lower coverage with the same number of attempts.

Due to the random nature of Monkey, not all events executed on the device will hit an interactive view in the application. Despite sending 500 events to the device, Monkey on average has only an 8\% hit rate. Monkey also takes approximately 5 minutes to execute. Given this time scale and the amount of interactions required to achieve a high coverage, larger applications cannot afford the wasted time and resources.

In comparison, Monkey++ achieves 100\% interface coverage with a maximum of 19 interactions. Knowing the underlying structure of the interface, removes redundant interactions and therefore saves time and resources. While gaining this knowledge requires pre-processing, it is only required once and the overhead is far less with our graph generation taking approximately 10 seconds on the largest AUT.

\section{Conclusion}
\label{Section:Conclusion}
Android's event driven paradigm and its reliance on callback methods has presented many challenges to traditional testing techniques. The most popular framework for testing is Exerciser Monkey, an automated random input generation tool designed to test an Android application through its interface. We argue that Monkey is a reliable, but limited, testing tool, that can be made more efficient and effective by replacing its random nature with knowledge of the applications structure.

In this paper, we presented a novel control flow structure able to represent the code of Android applications, including all interactive elements. We show that the (control flow) graph generated by our solution can direct/support the execution of a tool like Monkey (We call our extension \emph{Monkey++}) and address two main challenges in testing Android apps using the framework: removing duplicate/redundant interactions (i.e., efficiency, using less resources) by tracking areas already explored by the test suite; and increasing test coverage (i.e., its effectiveness) by targeting unexplored areas of the application.

Our results show that although Monkey can achieve a high coverage, it requires a large number of wasted interactions. Monkey achieves a coverage value of at least 85\%. However, to achieve these results, 500 interactions were executed and on average only 8\% actually interacted with the application. In comparison the test suite generated by Monkey++ achieved 100\% interaction coverage with a maximum of 19 interactions, showing that Monkey can be more effective with knowledge of the applications underlying structure.

In the future we will explore other search techniques suited to large scale Android applications while also leveraging knowledge of the applications internal structure (i.e., our control flow structure) to increase bug detection and efficiency.

\bibliographystyle{IEEEtran}
\bibliography{bibliography}

\end{document}